\documentclass[review]{elsarticle}

\usepackage{lineno,hyperref}

\journal{Journal of \LaTeX\ Templates}

\biboptions{authoryear}

\begin{document}

\begin{frontmatter}

\title{Penetrating particle ANalyzer (PAN)}





\author[mymainaddress]{X. Wu\corref{mycorrespondingauthor}}
\cortext[mycorrespondingauthor]{Corresponding author}
\ead{xin.wu@unige.ch}

\author[mysecondaryaddress2]{G. Ambrosi}
\author[mymainaddress]{P. Azzarello}
\author[mysecondaryaddress3]{B. Bergmann}
\author[mysecondaryaddress2,mysecondaryaddress4]{B. Bertucci}
\author[mymainaddress]{F. Cadoux}
\author[mysecondaryaddress5]{M. Campbell}
\author[mysecondaryaddress2]{M. Duranti}
\author[mysecondaryaddress2]{M. Ionica}
\author[mymainaddress]{M. Kole}
\author[mysecondaryaddress6,mysecondaryaddress7]{S. Krucker}
\author[mysecondaryaddress8]{G. Maehlum}
\author[mysecondaryaddress8]{D. Meier}
\author[mymainaddress]{M. Paniccia}
\author[mysecondaryaddress9]{L. Pinsky}
\author[mysecondaryaddress10]{C. Plainaki}
\author[mysecondaryaddress3]{S. Pospisil}
\author[mysecondaryaddress8]{T. Stein}
\author[mysecondaryaddress5]{P. A. Thonet}
\author[mysecondaryaddress2,mysecondaryaddress4]{N. Tomassetti}
\author[mymainaddress]{A. Tykhonov}

\address[mymainaddress]{Department of Nuclear and Particle Physics, University of Geneva, CH-1211, Geneva, Switzerland}
\address[mysecondaryaddress2]{Istituto Nazionale di Fisica Nucleare, Sezione di Perugia, I-06123 Perugia, Italy}
\address[mysecondaryaddress3]{Institute of Experimental and Applied Physics, Czech Technical University in Prague, 12800 Prague, Czech Republic}
\address[mysecondaryaddress4]{Dipartimento di Fisica e Geologia, Universit\`{a} degli Studi di Perugia, I-06123 Perugia, Italy}
\address[mysecondaryaddress5]{European Organization for Nuclear Research (CERN), CH-1211, Geneva, Switzerland}
\address[mysecondaryaddress6]{School of Engineering, University of Applied Sciences and Arts Northwestern Switzerland, CH-5210, Windisch, Switzerland}
\address[mysecondaryaddress7]{Space Sciences Laboratory, University of California at Berkeley, CA 94720, USA}
\address[mysecondaryaddress8]{Integrated Detector Electronics AS (IDEAS), NO-0484 Oslo, Norway}
\address[mysecondaryaddress9]{Physics Department, University of Houston, Houston, TX 77204, USA}
\address[mysecondaryaddress10]{Agenzia Spaziale Italiana, I-00133, Roma, Italy}

\begin{abstract}
PAN is a scientific instrument suitable for deep space and interplanetary missions. It can precisely measure and monitor the flux, composition, and direction of highly penetrating particles ($> \sim$100 MeV/nucleon) in deep space, over at least one full solar cycle (~11 years). The science program of PAN is multi- and cross-disciplinary, covering cosmic ray physics, solar physics, space weather and space travel. PAN will fill an observation gap of galactic cosmic rays in the GeV region, and provide precise information of the spectrum, composition and emission time of energetic particle originated from the Sun. The precise measurement and monitoring of the energetic particles is also a unique contribution to space weather studies. PAN will map the flux and composition of penetrating particles, which cannot be shielded effectively, precisely and continuously, providing valuable input for the assessment of the related health risk, and for the development of an adequate mitigation strategy. PAN has the potential to become a standard on-board instrument for deep space human travel. 

PAN is based on the proven detection principle of a magnetic spectrometer, but with novel layout and detection concept. It will adopt advanced particle detection technologies and industrial processes optimized for deep space application. The device will require limited mass (~20 kg) and power (~20 W) budget. Dipole magnet sectors built from high field permanent magnet Halbach arrays, instrumented in a modular fashion with high resolution silicon strip detectors, allow to reach an energy resolution better than 10\% for nuclei from H to Fe at 1 GeV/n. The charge of the particle, from 1 (proton) to 26 (Iron), can be determined by scintillating detectors and silicon strip detectors, with readout ASICs of large dynamic range. Silicon pixel detectors used in a low power setting will maintain the detection capabilities for even the strongest solar events. A fast scintillator with silicon photomultiplier (SiPM) readout will provide timing information to determine the entering direction of the particle, as well as a high rate particle counter. Low noise, low power and high density ASIC will be developed to satisfy the stringent requirement of the position resolution and the power consumption of the tracker.
\end{abstract}

\begin{keyword}
\texttt cosmic ray \sep solar energetic particle \sep space weather \sep space radiation \sep space travel \sep magnetic spectrometer \sep silicon tracker
\end{keyword}

\end{frontmatter}


\section{Introduction}

PAN, for Penetrating particle ANalyser, is an energetic particle detector for deep space application based on an innovative concept, aiming to make ground-breaking measurements crucial for space science and interplanetary exploration. It is a scientific instrument based on innovative concepts and technologies to fill an observational gap of energetic particles in deep space, outside the Earth's magnetosphere. PAN is a generic instrument specifically designed to be deployed on any deep space and planetary mission, even on future commercial interplanetary expeditions, with relatively low weight (20 kg) and low power consumption (20 W).  One possible mission opportunity is the Lunar Orbital Platform-Gateway (LOP-G)\footnote{\url{https://www.nasaspaceflight.com/tag/lop-g/}}
, the lunar-orbit station being developed by NASA and international partners.  

PAN will, for the first time, precisely measure and monitor over at least one full solar cycle (~11 years) the spectra, composition, and incoming direction of highly penetrating particles ($>\sim$100 MeV/nucleon) in deep space. Previous and current instruments cannot perform these measurements either because they are not operating in deep space, e.g. PAMELA ~\citep{PAM} and AMS-02~\citep{AMS}, or because the technologies used are not capable to precisely measure particle energies above a few 100 MeV/n, e.g. SOHO\footnote{\url{https://sohowww.nascom.nasa.gov/}}
, ACE~\citep{ACE} and the GOES\footnote{\url{https://www.nesdis.noaa.gov/GOES-R-Series-Satellites/}}
series of satellites.   
The science of PAN is multi- and cross-disciplinary, covering cosmic ray physics, solar physics, space weather and human space travel. PAN provides the enabling technology that will lead to ground-breaking measurements in these areas: 
\begin{itemize}
\item Perform, for the first time in deep space, precise flux and composition measurements of Galactic Cosmic Rays (GCR) in the 100~MeV/n to 20~GeV/n range, to further investigate the origin of these populations, and their propagation in the Galaxy and in the solar system.
\item Monitor, over at least one full solar cycle ($\sim$11 years), the GCR properties and correlate their variations with other solar observations, in order to investigate, in detail, the interplay between solar activity and the GCR propagation. 
\item Provide the first precise measurements of the flux, composition and arrival time at 1 AU (Astronomical Unit), outside the Earth's magnetosphere, of Solar Energetic Particles (SEP) in the 100~MeV/n -- 20~GeV/n range, during relativistic SEP events, associated with solar flares and Coronal Mass Ejections (CMEs). These measurements will allow to better determine the SEP energy spectrum during a specific event and, subsequently, to better understand the physical processes leading to SEP production and transport in the interplanetary space. The PAN measurements, therefore, are of paramount importance for determining both the detailed properties (e.g. spatial and energy distribution) of SEPs during different space weather conditions, as well as the acceleration mechanisms responsible for their generation. In this context, PAN will be an important asset for circum-terrestrial space weather monitoring, providing at the same time unique observational feedback for in-depth scientific research in Solar-Terrestrial physics.
\item By monitoring the deep space radiation environment continuously over a long period in a unique energy band outside the Earth's magnetosphere, with the possibility of pointing to any direction, PAN will be able to function as a space environment monitoring asset providing important information on the charged particle populations in the near-Earth space and long-term variability of their properties. The PAN data, therefore, will be a significant add-on for the development of global space weather models for long-term forecasting, based on the multi-wavelength and multi-messenger observations approach, in the context of both circum-terrestrial and planetary space weather. Moreover, understanding space conditions "at large" is a concept that is missing from current space weather models. 
\item Particles above 100 MeV/n cannot be shielded effectively thus posing serious health risks for human planetary exploration. PAN can provide real-time flux information that can be correlated with other in-situ radiation protection measurements to assess the health risks, and to develop mitigation strategies. Once its performance will be validated, PAN has the potential to become a standard on-board instrument for deep space manned missions. 
\end{itemize}

\section{Science objectives}
Cosmic rays, or energetic particles coming from the outer space, were discovered more than one hundred years ago (e.g. Hess' balloon flight in 1912). The study of cosmic rays has resulted in many discoveries in fundamental physics and space science (e.g. Anderson's discovery of the positron in 1932), and more recently, have contributed significantly to the understanding of the radiation environment near Earth and in the interplanetary deep space (e.g. AMS-02, SOHO, ACE, Voyager 1 and 2). Today, it is well established that space particle radiation comes mainly from 3 sources: particles trapped in the geomagnetic field, particles originating from the Sun, referred to as Solar Energetic Particles (SEPs), and Galactic Comics Rays (GCRs), the precise origin of which is still unknown. In deep space, the trapped particle contribution can be neglected, leaving only the SEPs and the GCRs as main contributors. 

In the absence of SEP events, GCR radiation is the dominant particle source above a few hundred MeV, consisting of mainly protons and Helium nuclei, but also heavier nuclei produced in stellar nucleosynthesis and through the interactions of GCRs with the interstellar medium.  In deep space the GCR flux below a rigidity\footnote{The (particle) magnetic rigidity is defined as the momentum of a charged particle divided by its charge: R = pc/Ze, in unit of GV, if the momentum unit is GeV/c. Particles with different properties (e.g. proton, He nuclei, etc.) will trace out the same trajectories in a static magnetic field provided that they have the same magnetic rigidity.} of ~20 GV is strongly affected by solar activity through the solar wind.

\begin{figure}[!htbp]
\begin{center}
\begin{tabular}{ll}
\includegraphics[width=0.4\textwidth]{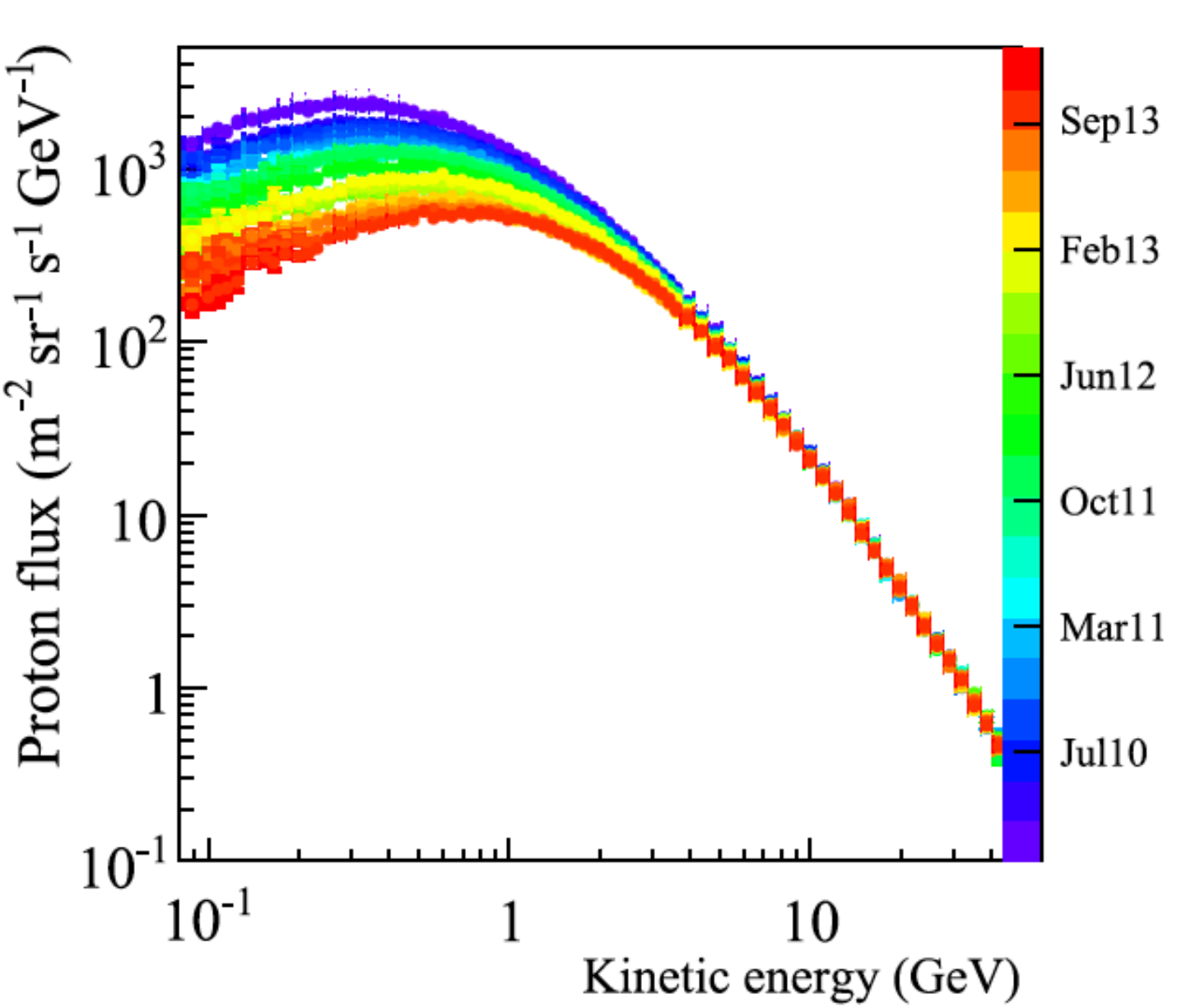} &
\includegraphics[width=0.5\textwidth]{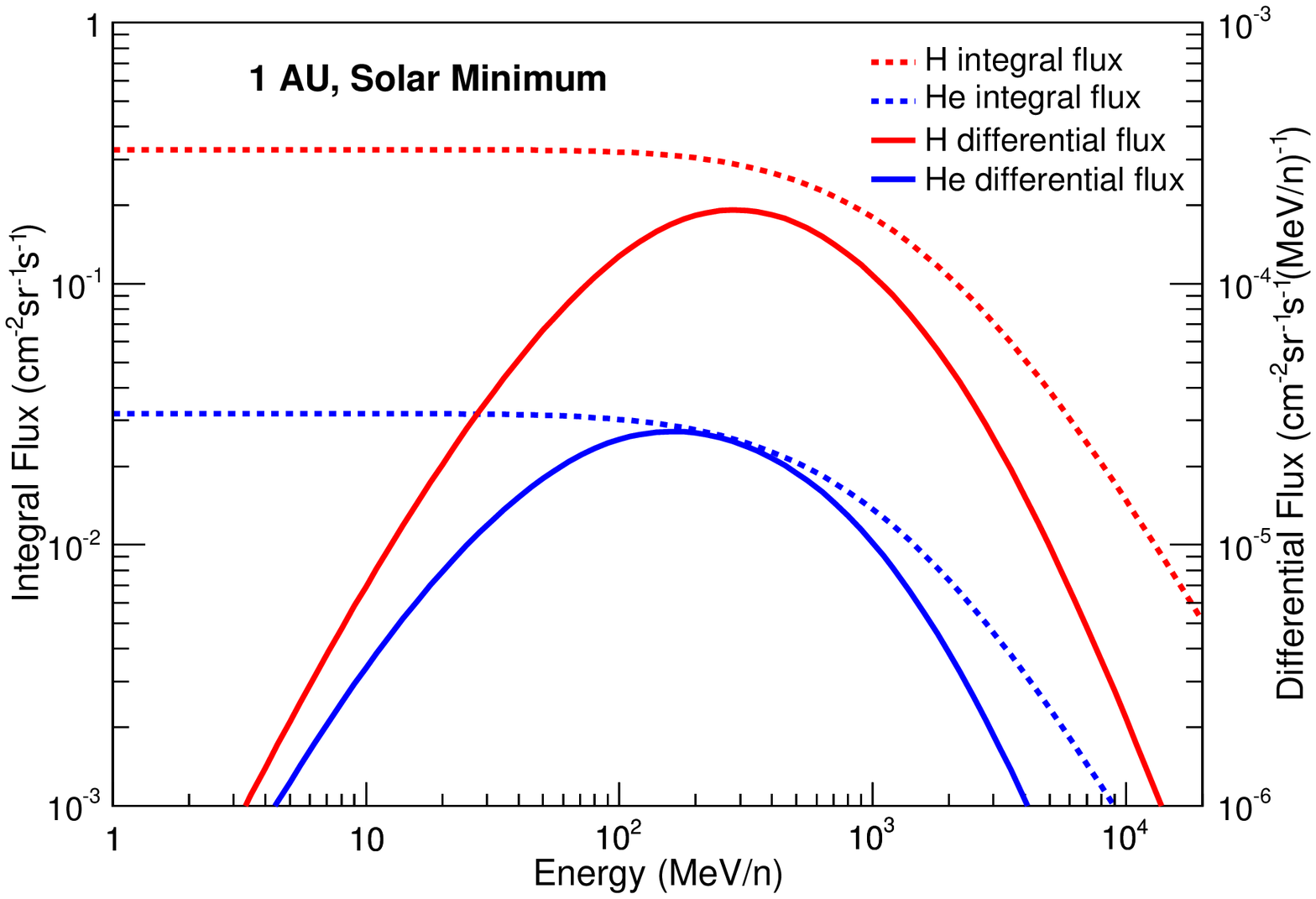} 
\end{tabular}
\end{center}
\caption{\small{(Left) The evolution of the proton spectra from the minimum to the maximum activity of solar cycle 24, from 2010 January (blue) to 2014 February (red) measured by PAMELA\citep{PAM_flux} near Earth. (Right) The integral and differential proton and Helium nuclei fluxes at 1 AU outside the Earth's Magnetosphere estimated with the SPENVIS toolkit.}} 
\label{fig:H_He_flux}
\end{figure}

The energetic particle environment at the Low Earth Orbit (LEO) has been studied in great detail, in particular since the launch of the PAMELA satellite in 2006 and the installation of the AMS-02 spectrometer on the ISS in 2011. The left panel of Figure~\ref{fig:H_He_flux} shows the evolution of the proton spectra from the minimum to the maximum activity of solar cycle 24,  measured by PAMELA in the near Earth space (similar measurements covering a more extended period, but with a higher rigidity threshold was reported by AMS~\citep{AMS_flux}. In deep space, however, only particles with energies below a few 100 MeV/n have been precisely measured; indeed, this low-energy particle component has been measured continuously over more than 17 years by the CRIS instrument on NASA's ACE mission. In addition, the HET instrument on Voyager 1~\citep{Voyager} has been sampling the GCR flux across the heliosphere in an energy range up to 500 MeV/n. While the particle flux above 20 GeV/n, predominantly from GCR, is largely unaffected by the geomagnetic field and the solar modulation, thus allowing the PAMELA and AMS measurements to be extrapolated into deep space, the same is not true for particles below this 20 GeV/n threshold. A precise measurement of particle flux between ~100 MeV/n to ~20 GeV/n in deep space is therefore missing (for a review of the current and planned particle observation missions, see e.g. reference~\citep{Laitinen}). PAN is designed to fill this gap, allowing to measure the particle flux in deep space in this energy range with unprecedented precision in terms of energy, composition, as well as short and long-term time variations. As shown in Figure~\ref{fig:H_He_flux} (Right), the total flux of GCR protons expected at 1 AU (Astronomical Unit) in deep space is about 0.2--0.3 particle/(cm$^2$sr$\cdot$s), estimated with the ESA's SPENVIS\footnote{\url{https://www.spenvis.oma.be/credits.php}}
toolkit. For an instrument with a geometrical acceptance (or Geometrical Factor, GF) of 10 cm$^2$sr, a rate of a few Hz is expected, which is sufficient for the GCR studies. Therefore, the design GF value of PAN is chosen to be 10 cm$^2$sr.

SEPs are bursts of energetic particles consisting mainly of electrons, protons and some heavier ions (e.g. He - Fe) with energies ranging from a few tens of keV/n up to several GeV/n~\citep{Malandraki}. SEPs are believed to be accelerated at the flare reconnection sites or/and at shocks driven by CMEs propagating through the solar corona and in the interplanetary space. Relativistic SEPs in the Earth's atmosphere can produce showers of secondaries with sufficient energy to be detected by ground-based neutron monitors, generating Ground Level Enhancement (GLE) events (e.g. the event on January 20 2005~\citep{Plainaki2007}). GLE events, therefore, are the manifestation of very intense solar activity requiring acceleration processes that produce SEPs with energies $\ge$500 MeV/n upon entry in the Earth's atmosphere~\citep{Plainaki}. One open science question, concerning the physics of SEP events, is the nature of the particle acceleration process. Several sophisticated mechanisms have been invoked~\citep{Aschwanden, Giacalone, Reames}, based often on the observational characteristics of the resulting SEP populations, to explain this acceleration such as the resonant wave-particle interactions occurring in conjunction with magnetic reconnection in the flare region, first-order Fermi acceleration, and shock-drift acceleration. However, the SEP flux profile is the superposition of particles accelerated at different locations (e.g. flare region and interplanetary space). Therefore, it is often difficult to disentangle the role of each acceleration mechanism and its relative efficiency, especially due to the fact that the available observations are often limited to the lower part of the energy spectrum (e.g. below GeV/n). As a result, at present, the principal acceleration mechanisms for SEPs are still under debate. PAN is designed to provide the much needed new input to improve the current SEP solar physics models and to revisit the proposed SEP acceleration scenarios.

SEP events are an important feature of solar-terrestrial physics and space weather; they have been closely monitored directly in space, for example currently by SOHO/EPHIN~\citep{Mellin}, ACE/CRIS and GOES-16/SEISS\footnote{\url{https://www.goes-r.gov/spacesegment/seiss.html}}
. However, energetic particles above ~100 MeV/n have not been precisely measured outside the Earth's magnetosphere, and most of the time only integrated flux is available. PAN will remedy this situation by providing the technology to measure continuously, precisely and in all directions (in scanning mode) these particles.

High-energy protons during the most intense SEP events can pose significant radiation hazards for astronauts and technological systems in space~\citep{Malandraki, Xapsos}. Most of the radiation risk for humans in space due to SEPs comes from proton energies above 50 MeV, the energy at which protons begin to penetrate spacesuits and spacecraft housing~\citep{Malandraki}. Moreover, SEP events with significant fluxes in the energy range above 100 MeV can be a severe radiation hazard to astronauts~\citep{Reames2013}. Therefore, knowledge of the SEP energy spectrum during such intense space weather events is essential. PAN will have the capability to measure the particle flux during rare but intensive relativistic SEP events (with particle energies up to a few GeVs) to provide critical input for scientific research in solar-terrestrial and space weather physics, with particular emphasis on the determination of the expected radiation environment over a long time period.  

In general, energetic particles above about 50--100 MeV/n, in particular protons and nuclei, cannot be shielded effectively, thus become "penetrating". These particles either pass the shielding material without losing much of their energy, or interact deep inside the material, generating a shower of secondary particles that can reach the protected enclosure. At LEO, the geomagnetic field is a natural shield for particles up to 20 GV in rigidity. In deep space however, without the geomagnetic shield, penetrating particles represent a serious radiation hazard for long term space travellers, even with the low flux GCRs. The precise and long-term measurements of the flux and composition of these particles are indispensable for the assessment of the related health risk, and the development of an adequate mitigation strategy. Moreover, the PAN measurements will provide an excellent opportunity to obtain further insights into the SEP spatial and energy distribution and to assess their effects on planetary space weather, an emerging research discipline (~\citep{Lilensten}, ~\citep{Plainaki2016}). In this context, it would be important to have particle detectors such as PAN, embarked in future Solar System missions as a basic payload requirement. PAN, if installed on a manned mission, such as the LOP-G, can provide valuable experience for the crew to operate a real-time penetrating particle monitoring tool.    

The ability of PAN to distinguish positively and negatively charged particles opens up very interesting opportunities in several research topics. For example, knowledge of the Interstellar Spectrum (IS) of GCR positrons is essential to understand the origin of these particles. In the low-energy range of $\sim$100 MeV, uncertainties in the positron IS amount up to one order of magnitude and the flux is subjected to large time-variation~\citep{Potgieter1, Potgieter2}. Moreover, the existing models of solar modulation encounter problems in describing the fluxes of leptons and hadrons under a self-consistent physical picture. In this respect, the PAN data will provide a clean, calibrated, and charge-sign resolved reference for the e$^+$/e$^-$ fluxes in the inner heliosphere (at $\sim$1 AU). Also, in both direct and indirect search of dark matter particles, several recent efforts are being oriented at particles masses in the $\sim$50--1000 MeV/c$^2$ scale~\citep{Alexander}. Indirect searches, however, suffer from lack of antimatter GCR data at these energies: the best upper limits on the annihilation rate have been found with the Voyager-1 data on the total electron and positron flux (i.e. charge-sign unresolved) collected in the interstellar space~\citep{Boudaud}. PAN can provide precise GCR antiparticle measurements, allowing to reach highly competitive limits over an extended dark matter mass range, with strengths at the level of those obtained from the CMB.

\section{Instrument concept and measurement principle}
PAN is based on the particle detection principle of a magnetic spectrometer, with novel layout and detection concepts to optimize the measurement precision for both high flux and low flux particles. It adapts advanced particle detection technologies and industrial processes for their optimal application in deep space, pushing the technological limits in several key areas. It is designed to precisely measure the momentum, the charge, the direction and the time of galactic cosmic rays and solar energetic particles between 100~MeV/n and 20~GeV/n. It can also measure the integral flux of particles above 10 MeV/n.

The energy of a charged particle entering a magnetic spectrometer is derived from the degree of bending of the particle trajectory in the magnetic field, assuming that the particle identification is performed independently. This is the only practical way to measure the energy of nuclei in the GeV/n range to better than 20\%. In PAN, the bending of the particle in the magnetic field is measured by precise silicon strip tracking detectors, while the elemental identity of the particle is determined by its charge, Z, measured with the dE/dx method at multiple points: the tracking detectors, the time-of-flight (TOF) detectors, and the pixel detectors. 

The unique features of PAN, with respect to other spectrometer experiments (e.g. PAMELA and AMS) are: 
\begin{itemize}
\item	Both the bending radius and bending angle are measured, which, together with a segmented magnet system, allows to increase the geometrical acceptance and improve the energy resolution.
\item	The bending and the charge of the particle in the tracker are measured by different silicon strip detectors, which allow to independently optimize the detector and readout ASIC design for these two measurements.
\item	Very fine pitch silicon detectors are used to achieve the best possible position resolution in the bending direction, improving energy resolution at high energy.
\item	The thickness of the silicon detector is minimized to reduce multiple Coulomb scattering effect, improving energy resolution at low energy.
\item	Silicon pixel detectors in low power settings are used to enhance high rate capability. 
\item	The instrument is symmetric, effectively doubling the geometrical acceptance for isotopic particles such as GCRs.
\end{itemize}

\begin{figure}[!htbp]
\begin{center}
\includegraphics[scale=0.65]{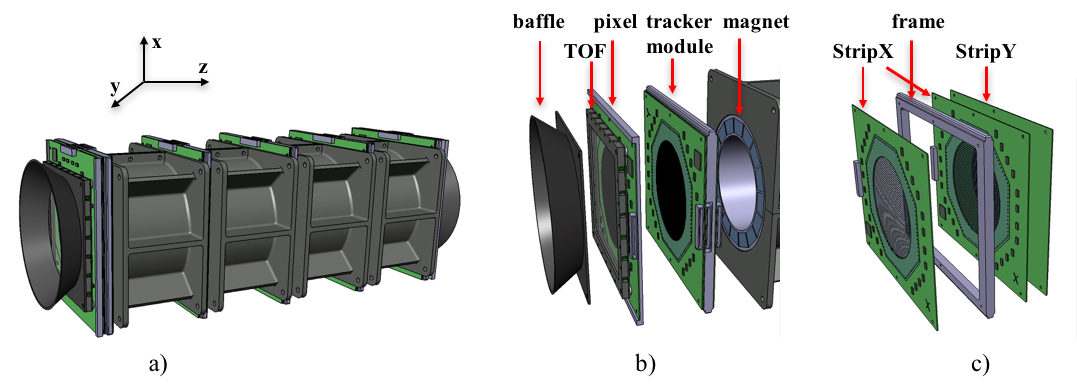}
\caption{\small{a) A drawing of the PAN instrument. b) Exploded view showing the baffle, the TOF module coupled to the Pixel module, the tracker module and part of a magnet sector. c) Exploded view of a tracker module showing the 2 StripX layers with a support frame in between, and the StripY layer after the second StripX layer.}}
\label{fig:PAN_sketch}
\end{center}
\end{figure}

These new features allow to produce a high impact space instrument with relative low mass and low power budget, suitable for deep space and interplanetary missions. The design concept of PAN is shown in Figure~\ref{fig:PAN_sketch}. It consists of a cylindrical magnetic spectrometer (MS), with a fast TOF detector coupled to a pixel detector at each end. The MS is about 44 cm long, segmented into 4 sectors and instrumented with 5 tracker modules, each consisting of 3 layers of silicon tracking detectors, two to measure the bending plane coordinate (StripX), one for the non-bending plane coordinate (StripY). Two StripX layers per module are implemented in order to measure the bending angle of a particle after having traversed a magnet sector. Immediately after the TOF detector is a silicon pixel detector (Pixel) in low power settings  with high rate capability, in order to maintain the detection capabilities for even the strongest solar events. The TOF and StripY detectors provide the trigger signal, as well as the timing information to determine the entry end of the particle. They are also capable of measuring charge of the particle up to Z = 26. 

The PAN instrument concept is robust, scalable and modular, with high redundancy for position, charge and time measurements. Measuring modules can be made mechanically independent from the magnet, allowing them to be exchanged in view of a possible extension of the instrument lifetime.  

The key functionalities and components and the associated challenges of the PAN instrument are described in more detail below.

\paragraph{Spectrometer (magnet and silicon tracker)} The dipole magnet sectors will be built from Halbach arrays of high field NdFeB permanent magnet, instrumented in a modular fashion with fine pitch silicon strip detectors in the bending plane, allowing to reach an energy resolution better than 10\% for nuclei from H to Fe at 1 GeV/n. The Halbach~\citep{Halbach} scheme is an elegant and flexible way to produce an enclosed and uniform magnetic field by arranging small pieces of permanent magnets according to a well-defined formula. The key parameters of the spectrometer are the magnetic field strength, the total length of the magnet, the thickness of the tracking detectors (StripX, StripY), and the position resolution of the silicon strip detector in the bending plane (StripX). The challenge is to use fine pitch (25 $\mu$m) and thin (150 $\mu$m) silicon strip detectors to reach a position resolution of 2 $\mu$m, which has not yet been achieved with long strips of 10 cm. The density of the readout channel (up to 4000 channels per 10 cm) is pushing the detector fabrication and assembly technology to their limits.

\paragraph{Particle identification} The charge of the particle can be determined by measuring the energy deposited in different layers of the detector (the dE/dx method): scintillating detectors (TOF), pixel detectors (Pixel) and silicon strip detectors (StripX, StripY). The multiple measurement points allow the full Z range, from Z = 1 to 26, to be optimally covered, reaching a resolution of 10-20\% of Z. The main challenge is the large dynamic range required for the readout system. In particular, the readout ASICs for the StripY detectors has to be able to provide a dynamic range of about 1$\times$10$^5$.

\paragraph{Pixel detector} It will be the first time that a large area active silicon pixel detector is used in space. The challenge is the power consumption due to the large channel count. PAN will profit from the latest advances on power management schemes of active pixel detector, in particular the Timepix\footnote{\url{https://medipix.web.cern.ch/}}
series detectors. Different power saving scenarios will be investigated. Using a pixel detector with PAN allows to maintain particle identification capabilities at very high rates, even during very strong solar events, which is very important for solar physics and space weather studies.   

\paragraph{Time-Of-Flight (TOF) detector} Fast plastic scintillators with silicon photomultiplier (SiPM) readout will provide timing information to determine the entering end of the particle. The TOF detector will measure the charge of the particles, provide a time of flight measurement with a precision of 100 ps, and provide a high rate (up to 10 MHz) counter for particles above 10 MeV, particularly important during solar events.

\paragraph{Readout ASICs}  Low noise, low power, high density, large dynamic range Application Specific Integrated Circuits (ASICs) need to be developed to satisfy the stringent performance requirements within the programmatic constraints. 

\section{Instrument design consideration and performance assessment}

\paragraph{Energy measurement} Magnetic spectrometer is a proven detection technology to measure the particle rigidity, thus energy, for energy between 100 MeV/n and 20 GeV/n (e.g. PAMELA and AMS-02). The classical $\Delta$E--E method used by many deep space instruments (e.g. the CRIS instrument on ACE) cannot be applied in this energy range since particles start to shower in the detector, so the energy resolution is unavoidably degraded by the showering process. In addition, a very thick, thus heavy, calorimeter would be needed to stop the particle to measure the total energy. The MS method, on the other hand, detects the bending of a charged particle in a magnetic field to measure its rigidity, from which the energy can be derived, if the identity of the nucleus is determined independently by its charge measured by the charge detectors. In this case the rigidity resolution, thus the energy resolution, has 2 contributions. One, called measurement effect, is related to the magnetic field (strength and length) and the tracker precision, and it increases with momentum. The other, due to the Multiple Coulomb scattering (MCS), is related to the thickness of the detector, and it decreases with the particle momentum. Thanks to the technological advance in particle detection technology in the last decade, reliable thin silicon detectors with very good position resolution are now commercially available.  With appropriate instrument design, it is then possible to mitigate the two effects to achieve a good energy resolution over the desired energy range, with a magnet of reasonable mass. 

\begin{figure}[!htbp]
\begin{center}
\includegraphics[scale=0.45]{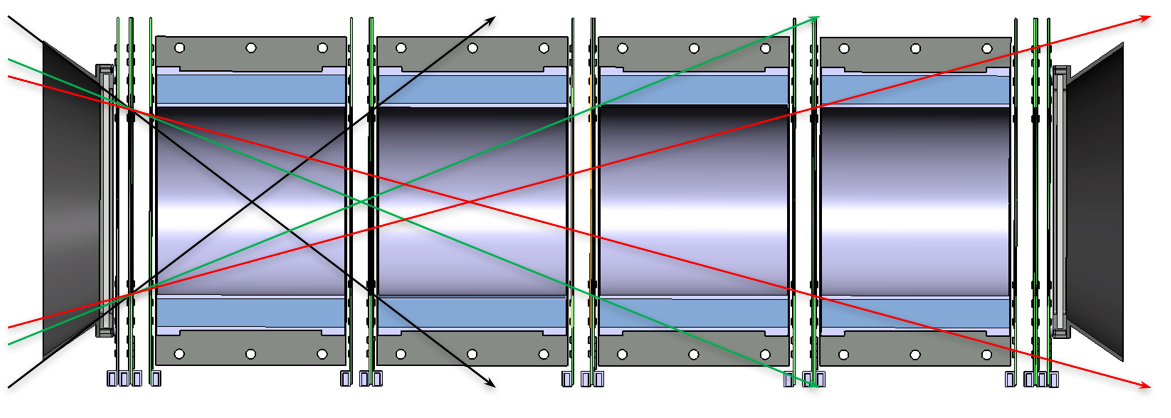}
\caption{\small{Illustration of different geometrical acceptance for particle traversing 1 (black), 2 (green) and 3 (red) magnetic sectors. }}
\label{fig:PAN_GF}
\end{center}
\end{figure}

The energy resolution of the spectrometer can be estimated with the Gluckstern formulas~\citep{Gluckstern}. Since the resolution depends on the path length of the particle in the magnetic field, particles traversing more magnet sectors will have better resolution, but with a reduced geometrical acceptance, as illustrated in Figure~\ref{fig:PAN_GF}.   

The design of PAN thus allows to compensate the worsening of the energy resolution with increasing statistics: for high flux particles full passage events can be used to profit from the best energy resolution, while for low flux events the measurement precision is dominated by statistics, so accepting large number of short passage events can bring significant improvement. The geometrical acceptance, also called geometrical factor (GF), of the baseline layout (4 magnet sectors of 10 cm in length and diameter, with a gap of 1 cm between sectors), for particles passing 4, 3, 2 and 1 magnet sectors, are about 6, 10, 21 and 65 cm$^2$sr respectively, taking into account that the instrument can detect particles arriving from both ends. The corresponding opening angles are about 25$^{\circ}$, 33$^{\circ}$, 47$^{\circ}$ and 80$^{\circ}$.

\begin{figure}[!htbp]
\begin{center}
\begin{tabular}{ll}
\includegraphics[width=0.5\textwidth]{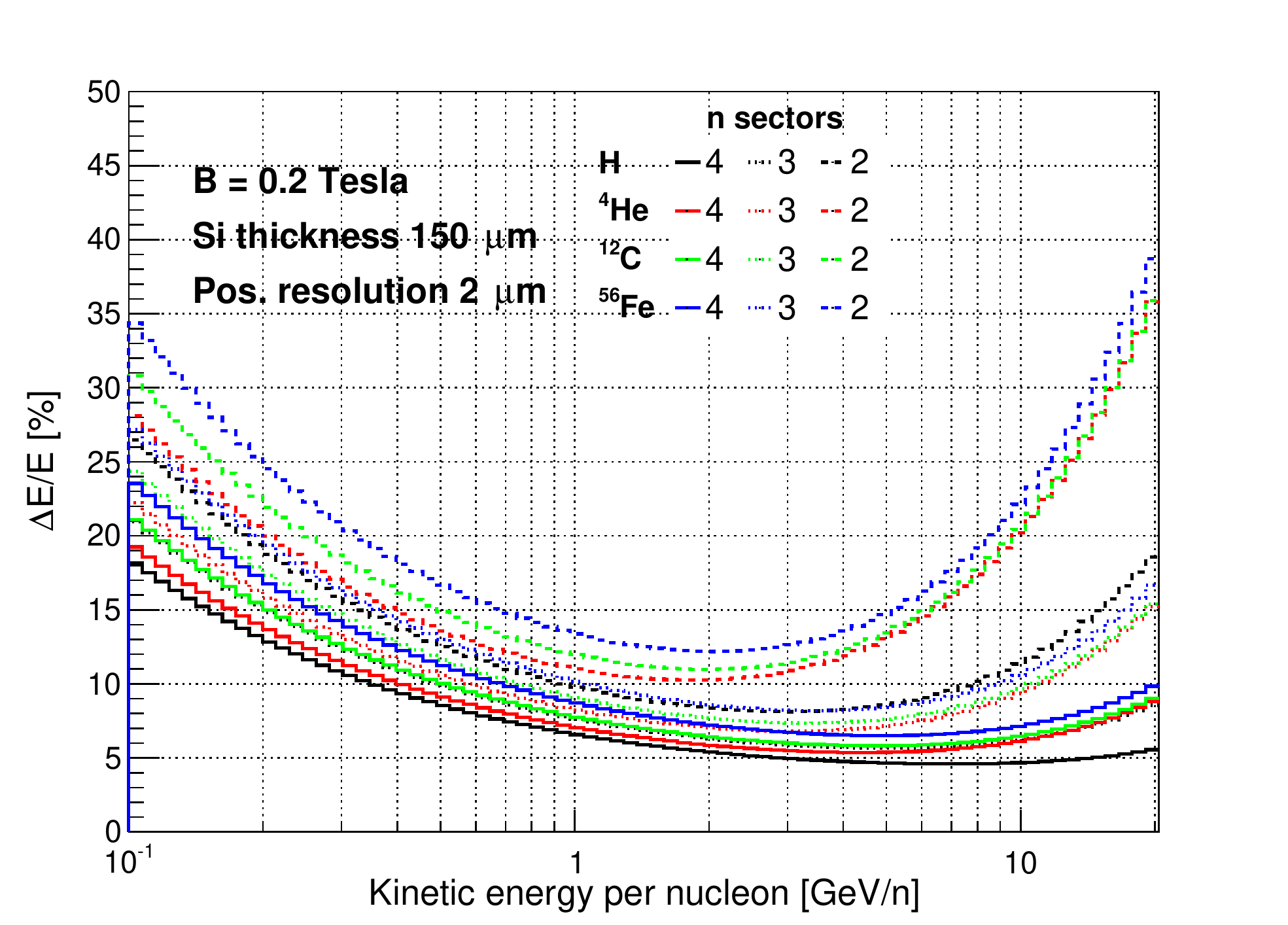} &
\includegraphics[width=0.5\textwidth]{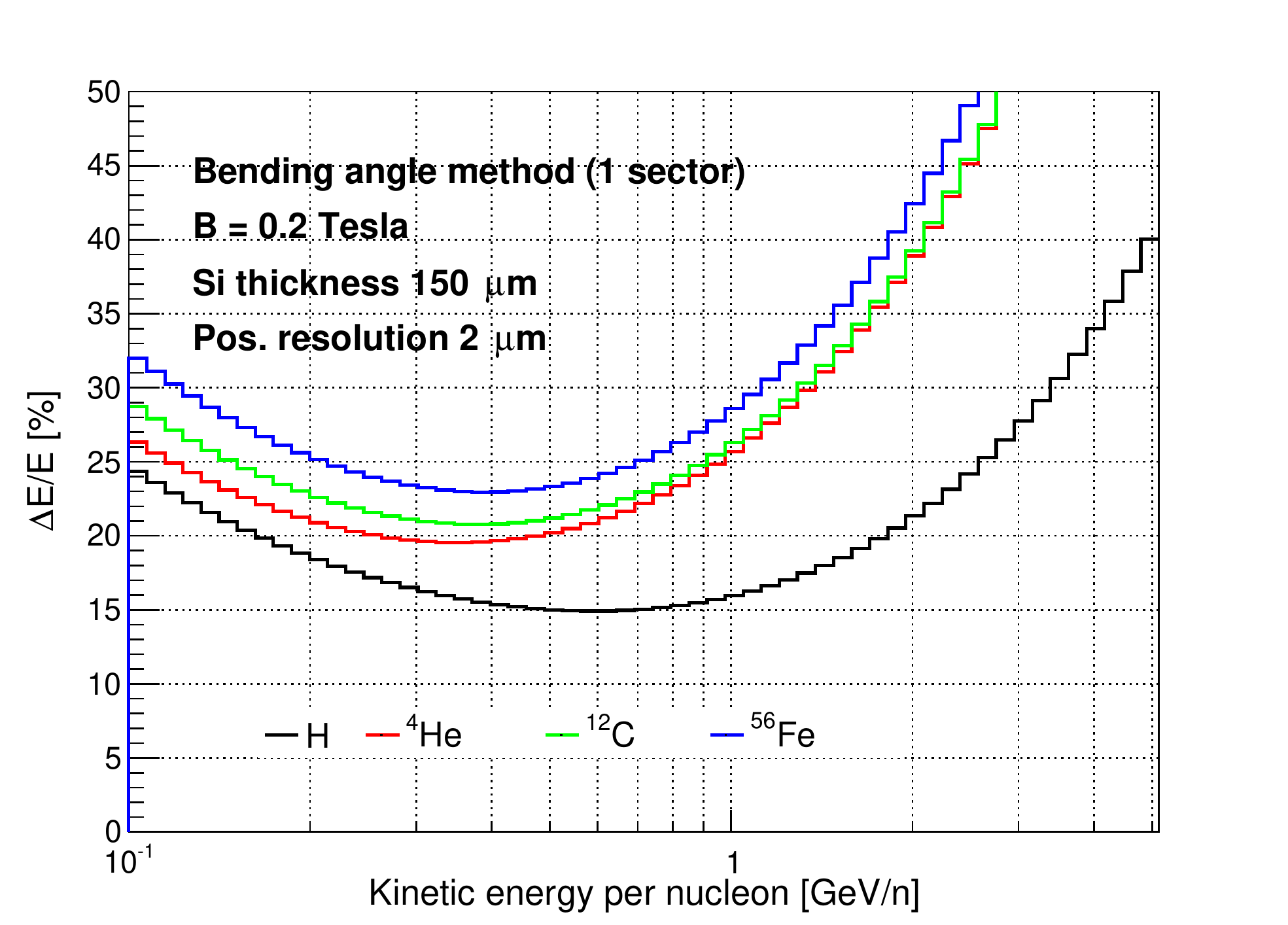} 
\end{tabular}
\end{center}
\caption{\small{(Left) Energy resolution as a function of energy per nucleon for proton, Helium, Carbon and Iron nuclei passing 4, 3, 2 magnet sectors, from 100 MeV/n to 20 GeV/n, with the baseline PAN design. (Right) Energy resolution as a function of energy per nucleon for proton, Helium, Carbon and Iron nuclei passing only 1 magnet sector, from 100 MeV/n to 10 GeV/n.}}
\label{fig:res}
\end{figure}

Figure~\ref{fig:res} (left) shows the energy resolution obtained with the baseline PAN design with realistic parameters: magnetic field B = 0.2 T, magnet sector length h=10 cm, bending plane position resolution $\delta$x = 2 $\mu$m, tracking detector thickness 150 $\mu$m, for particles passing 4, 3, 2 magnet sectors, for 4 types of nuclei (H, He, C, Fe). The requirement of a GF of 10 cm$^2$sr with resolution of 20\%(50\%) for H(Fe) between 100 MeV/n to 20 GeV/n can be satisfied by events passing the 3 sectors. In fact, for particles passing 2 magnet sectors, which corresponds to a GF of 21 cm$^2$sr, the energy resolution is still very good (better than 30\% and 40\% for proton and Fe, respectively, in the energy range of 100 MeV/n to 20 GeV/n).

Even particles traversing only 1 segment can be used since the momentum can be measured by the bending angle method, which will have an opening angle of 80$^{\circ}$ and GF=2$\times$32.5 cm$^2$sr. This is the reason for the two x-layer (StripX) at each measuring station (tracker module) in Figure 2. With a nominal distance of 10 mm between the two StripX layers, the energy resolution for nuclei is ~25-50\% for $<$ 1 GeV/n, as shown in the right plot of Figure~\ref{fig:res}.   

Note that the tracker is self-aligned, with a few thousand particles passing the full length of the instrument. This requires the temperature of the instrument to be stable within 1-2 $^{\circ}$C per hour.

\begin{figure}[!htbp]
\begin{center}
\begin{tabular}{ll}
\includegraphics[width=0.5\textwidth] {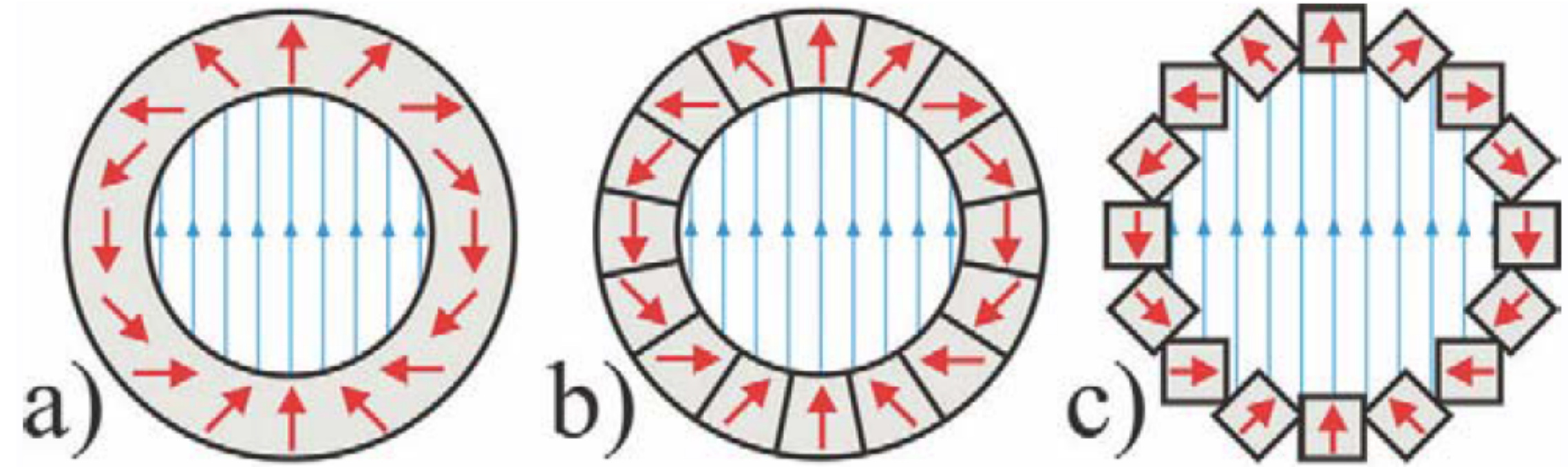} &
\includegraphics[width=0.35\textwidth]{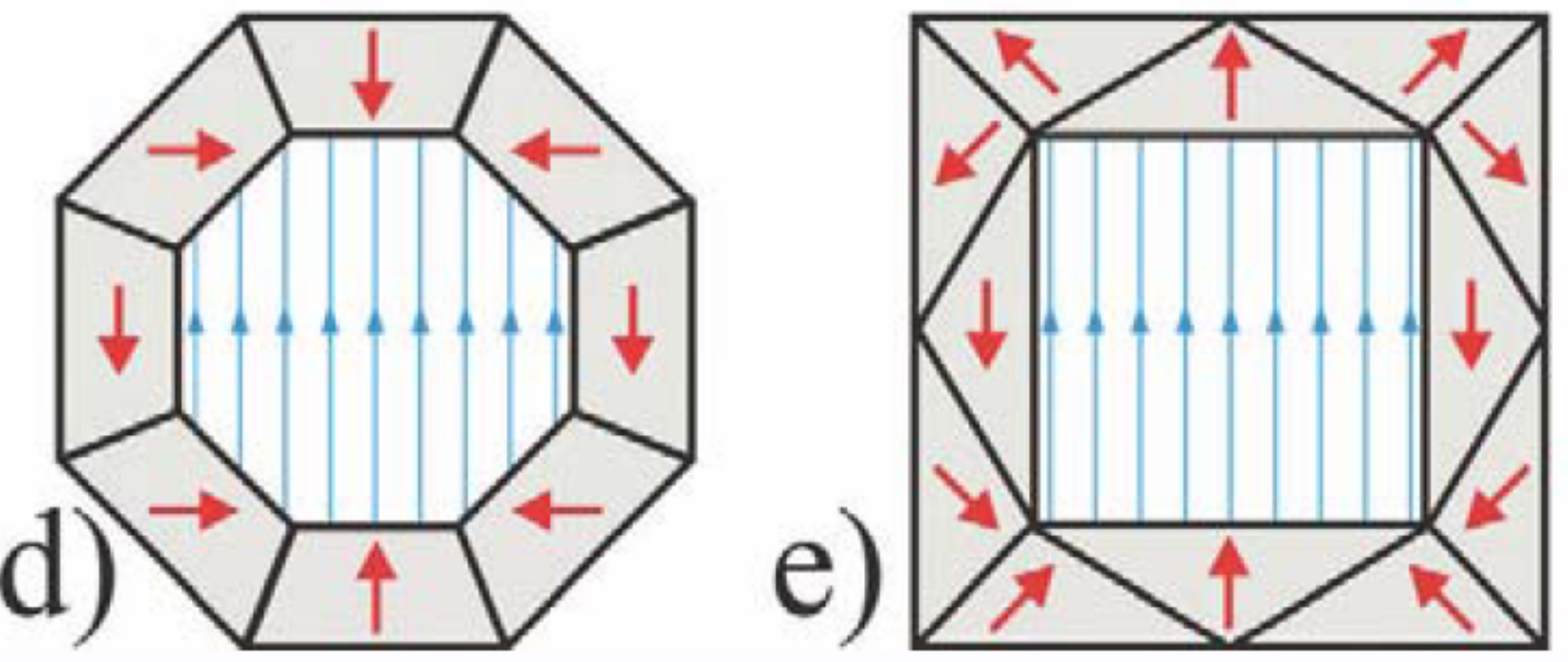} 
\end{tabular}
\end{center}
\caption{\small{(Left) Different layouts of Halbach dipole magnets (b), (c) (d), (e), compared to the ideal magnet (a), from reference~\citep{HAL_scheme}.}} 
\label{fig:hal}
\end{figure}

\paragraph{Magnet system}  The magnet system to a large extent defines the performance and the weight of the instrument. For robustness and accessibility, the PAN magnet is segmented into four magnetic sectors, each is a cylinder of 10 cm inner diameter and 10 cm long, for a total length of 40 cm. The magnetic field is generated by arranging blocks of NdFeB permanent magnet according to the Halbach scheme to produce a dipole magnetic field. Halbach array magnets have been successfully used in particle accelerators~\citep{Thonet} and Nuclear Magnetic Resonance (NMR) imaging~\citep{Soltner}. There are many possibilities to arrange Halbach arrays, as shown in Figure~\ref{fig:hal}. For the PAN baseline design, a layout that is very close to the ideal case (b in Figure~\ref{fig:hal}) will be used.  

The thickness of the magnet needed can be obtained by the formula $B=B_{r}F{\ln}(R/r)$, where $r$ and $R$ are the inner and outer radius of the cylinder, $B_r$ the remnant field of the NdFeB magnet, $B$ the resulting dipole field, and $F$ the filling factor which is a function of the number of magnet wedges ($M$) used to enclose the magnetic field circumference~\citep{Halbach1985}: $F=sin(2{\pi}/M)/(2{\pi}/M)$. Assume 16 wedges are used then $F = 0.974$. To generate a B field of 0.2 T with magnets of $Br = 1.4$~T, an $R = 5.79$~cm is needed according to the Halbach formula, so the thickness of the magnet is about 8 mm, which results in a total magnet weight of $\sim$8 kg. Extra mass would be needed to support and shield the magnet. With 1~mm Al shielding the Total Ionization Dose (TID) is expected to be $\sim$20~krad for 12 years, according to the SPENVIS simulation, which should not cause degradation of the magnet. 

\begin{figure}[!htbp]
\begin{center}
\begin{tabular}{ll}
\includegraphics[width=0.4\textwidth] {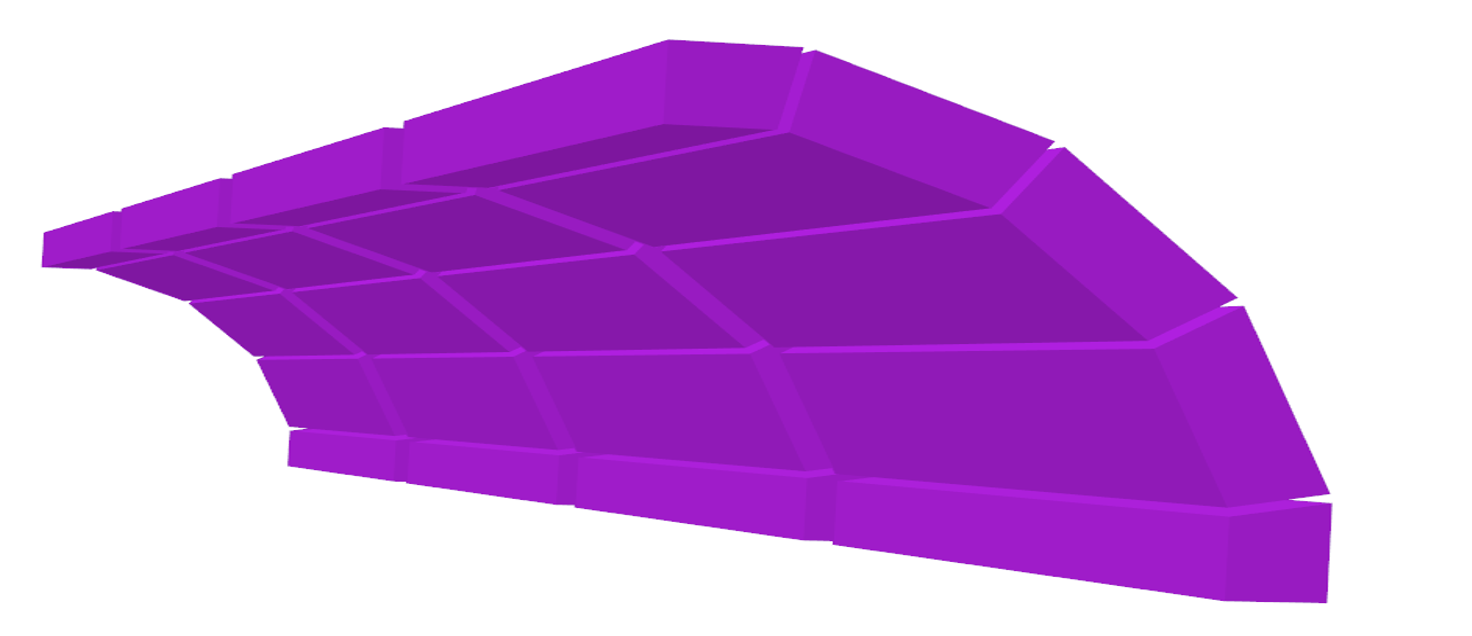} &
\includegraphics[width=0.55\textwidth]{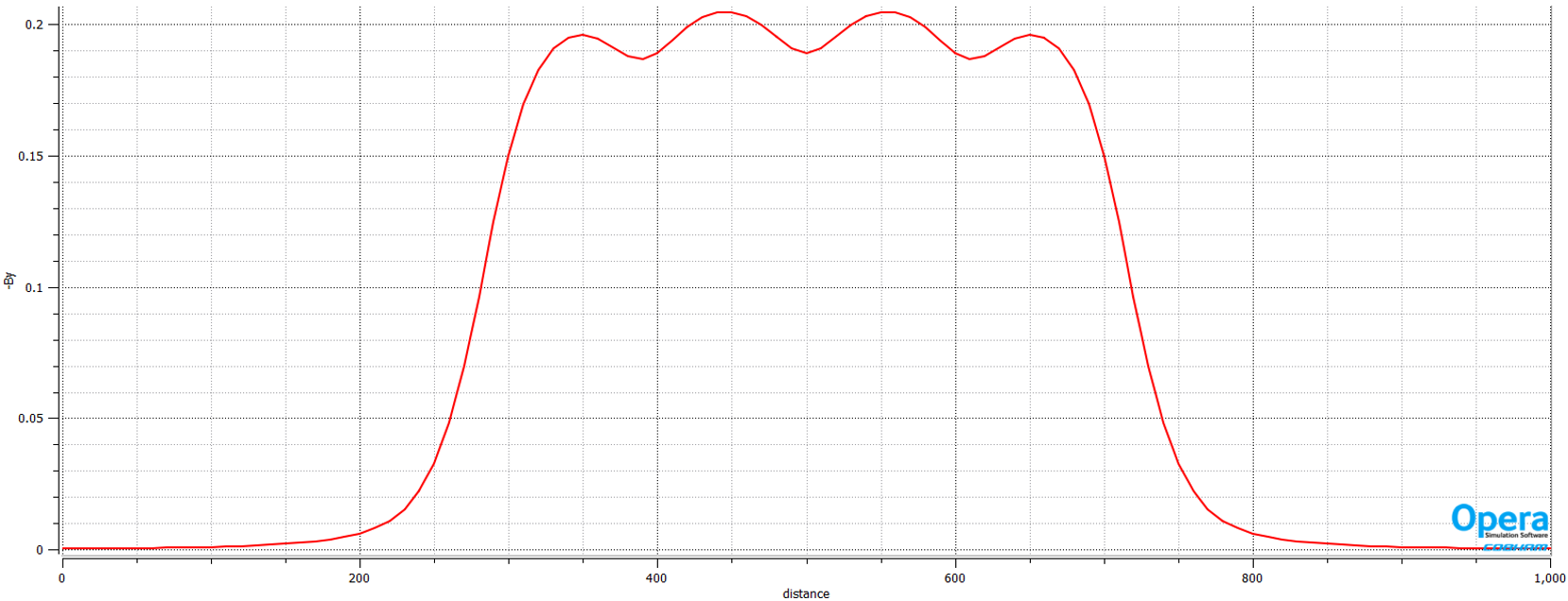} 
\end{tabular}
\end{center}
\caption{\small{(Left) 3D drawing of $1/4$ of the preliminary magnet design (left) , and the B-field along the axis of the magnet resulting from the simulation (right). Only $1/4$ is designed because of the dipole symmetry.}} 
\label{fig:Thonet}
\end{figure}

Figure~\ref{fig:Thonet} shows the 3D model of the preliminary design of the 4-sector magnet (left), and the simulated B-field produced along the axis  of the instrument (right), which shows a field uniformity within 5\% around the nominal 0.2~T.

\paragraph{Silicon Tracker} The bending of the charged particle trajectory (track) can be measured precisely by silicon strip detectors, a technology successfully applied by the PAMELA and AMS-02 missions. Both PAMELA and AMS-02 used double-sided detectors. In view of its long operation time ($>$~11 years) expected in deep space, and the stringent requirement of position resolution on the bending plane, the more robust single sided technology will be used. Also, using separate sensors to measure the bending plane (X) and non-bending plane (Y) coordinates provides more flexibility for the optimization. 

As shown in Figure~\ref{fig:PAN_sketch}, there are 5 tracker modules, each consisting of one StripY sensor and two StripX sensors. Two StripX sensors per module are needed to measure the bending angles between a magnet sector. The nominal distance between the 2 StripX sensors is 1 cm; the StripY sensor should be as close as possible to one of the StripX sensors.  

For the bending direction coordinate measurement, a fine pitch strip detector will be used. The smallest implant pitch of commercially available detectors is 25 $\mu$m. If every other strip is readout, thanks to the floating strip charge sharing mechanism~\citep{charge}, it is possible to reach a position resolution around 5  $\mu$m as achieved by several experiments, e.g. by PAMELA~\citep{PAMELA_tracker}. In order to have the best possible energy resolution, PAN will push this limit by developing a silicon strip detector with 25~$\mu$m readout pitch to reach a 2 $\mu$m resolution, because with such small pitch, the deposited charge is always shared between 2 strips, even without floating strip, so the position resolution is improved. However, this performance has never been achieved before for a detector with such a long strip (up to 10 cm), and such small thickness (150 $\mu$m). Preliminary discussion with a silicon detector manufacturer has confirmed the feasibility of such design, and an innovated layout scheme is being developed to meet the challenge of fanning out the 4000 strips over 10 cm on the sensor, to be bonded to the readout ASICs. The total capacitance of a 25 $\mu$m pitch strip of a 150 $\mu$m thick sensor is about 2-2.5 pF/cm according to simulation, well suited for low noise readout ASICs.  

Given the large channel count (40,000 for the full instrument) a stringent power budget of $<$0.2 mW/channel is required for the readout ASIC. With this specification the expected power consumption for the 10 StripX sensors in the instrument will be about 8 W. A fall-back solution is to use a 50 $\mu$m readout pitch, which still provides an energy resolution within the performance requirement. The power consumption will be reduced by 50\%, however the resolution for the particle passing only one magnet sector will be degraded significantly at high energy. 

The StripY sensors can have a much larger pitch because they are used to measure the particle pointing direction, which can profit from the large lever arm of 10 cm between two Y measurements. The baseline StripY sensor has a pitch of 500 $\mu$m, corresponding to a position resolution of ~144 $\mu$m, which gives an angular resolution of ~0.12$^\circ$, which would be more than adequate for the PAN application. The total number of channels per sensor is 200, and 1000 for all 5 sensors in the instrument. As will be explained below, StripY sensors are also used to measure the charge up to Z = 26, thus the power consumption per channel is higher, up to 1 mW, for a total power consumption of 1~W. 

\paragraph{Charge measurement} In PAN the charge of the particle (Z) is determined by measuring the energy deposited ($dE/dx$) by the particle in detector layers, since $dE/dx$ is proportional to Z$^2$ according to the Bethe-Bloch formula. However, $dE/dx$ is a quantity with large intrinsic fluctuation (described by the Landau distribution), it is essential to have multiple independent measurements, as demonstrated by the AMS-02 and PAMELA missions. In PAN, from 7 to 17 dE/dx measurements are available, depending on the geometry, with 4 types of detectors: the scintillator based TOF detector, the silicon pixel detector, and the silicon strip detectors (StripX/StripY). To measure Z up to 26, the readout system needs to have a large dynamic range (DNR), in particular the front-end ASICs, which will have a large impact on the power consumption. The strategy of PAN is to optimize the readout of the TOF and StripY for the large Z measurement, which has low rate, and the Pixel and StripX for low Z but at high rate. Multi-range readout is necessary for TOF and StripY ASICs. The silicon pixel detector can provide unambiguous charge identification in very high multiplicity events, particularly interesting for studying strong solar particle events. A promising candidate is the Timepix~\citep{TimepixPaper} series hybrid pixel detector, which has already been used in space for dosimetry measurements~\citep{Turecek} and space weather analyses~\citep{Gohl} . But it needs to be re-optimized for large dynamic range and low power consumption.

\begin{figure}[!htbp]
\begin{center}
\begin{tabular}{ll}
\includegraphics[width=0.5\textwidth] {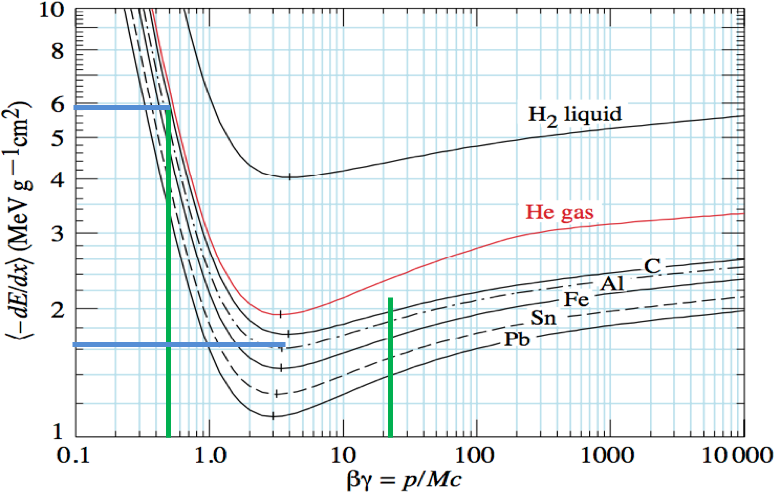} &
\includegraphics[width=0.35\textwidth]{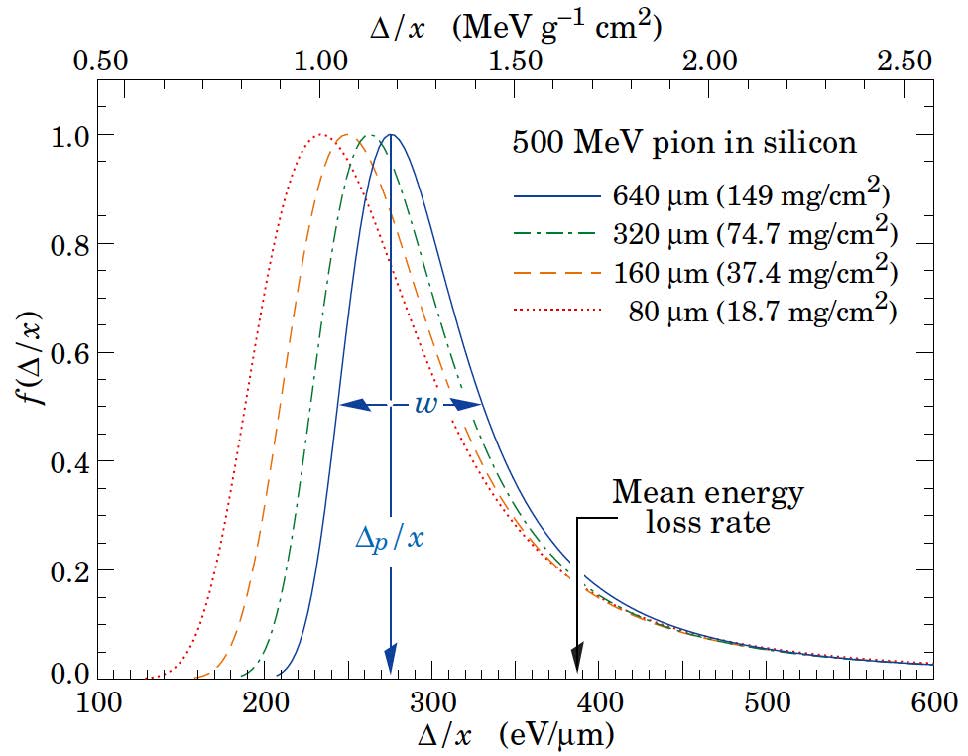} 
\end{tabular}
\end{center}
\caption{\small{Left) Mean energy loss (dE/dx) of Z = 1 particle in different material as a function of $\beta\gamma$. (Right) Distribution of energy loss in unit length of a MIP particle (pion of 500 MeV, $\beta\gamma$ = 4) in different thickness of silicon. From reference~\citep{PDG}.}}
\label{fig:dedx}
\end{figure}

The requirement on the DNR needs also to take into account the fact that at 100 MeV/n, the $\beta\gamma$ of the particle is about 0.5, which leads to a $dE/dx$ that is ~3.2 times of the Minimum Ionization Particles (MIP) at $\beta\gamma$ $\sim$4, as shown in Figure~\ref{fig:dedx} (left). 

As mentioned above $dE/dx$ follows a Landau distribution, as shown in Figure~\ref{fig:dedx} (right). For 150 $\mu$m silicon, the most probable value (MPV) for a Z=1 MIP particle is about 38 keV, which will generate 11k pairs of electron-hole, which, if all collected, corresponds to 1.7 fC of charge. For a Z = 26 particle the MPV is about 1.1 pC at MIP energy and 3.7 pC at 100 MeV/n. Taking into account the large tail of the Landau distribution (0.1-2 MPV), the system needs to measure the input charge up to 8 pC for the silicon strip detectors. The dynamic range from 0.085 fC (1/20 of MIP charge) to 8 pC is 1x10$^{5}$. Similar dynamic range is required for the TOF readout system.  Note that the identification of electrons is straightforward since they bend in the MS to a direction opposite to that of the nuclei, and their dE/dx vs. momentum behaviour is very different from that of nuclei. 

\paragraph{Time of Flight measurement and the TOF detector}. The TOF detector is the first detector encountered by particles entering the instrument. Thin window and Multi-Layer insulation (MLI) equivalent to 600 $\mu$m aluminium will block charged particles below 10 MeV/n from reaching the TOF, and a baffle will be used to limit the acceptance of the TOF to the maximal acceptance of the instrument, of 32 cm$^2$sr. The TOF detector will be used to:
\begin{itemize}
\item Measure the integral flux of particles above 10 MeV during solar events.
\item Provide a timing signal for the time-of-flight measurement to determine the entry end of the particle, together with the timing signal measured by StripY sensors in the spectrometer.
\item Measure the charge of the particle, from Z = 1 to 26.
\item Stop particles below 20 MeV from reaching the spectrometer.
\end{itemize}

According to the NOAA definition\footnote{\url{https://www.swpc.noaa.gov/noaa-scales-explanation}}
, an extreme Solar Radiation Storm, which occurs fewer than once per solar cycle, will have a flux of about 10$^5$ proton/(cm$^2$sr$\cdot$s) above 10 MeV, which corresponds to 3.2 MHz for a GF of 32 cm$^2$sr. Therefore, the TOF will be designed to be able to count particles up to 10 MHz.

\begin{figure}[!htbp]
\begin{center}
\includegraphics[scale=0.35]{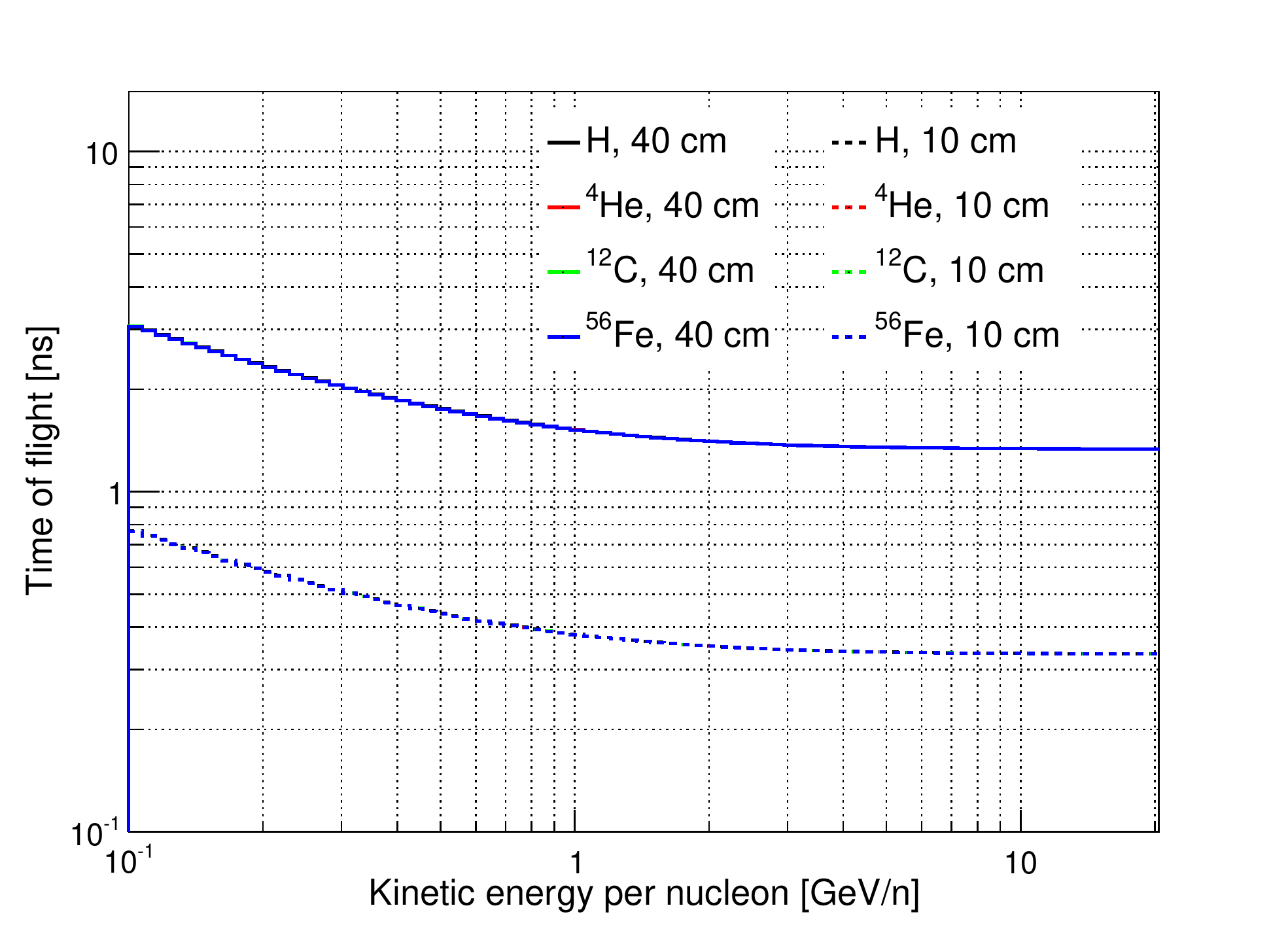}
\caption{\small{Time of flight of nuclei as a function of the kinetic energy per nucleon, for particles passing the full spectrometer length (40 cm) and one magnet sector (10 cm). The curves of different isotopes are very close to each other since their velocities are almost the same for a given kinetic energy per nucleon.  
}}
\label{fig:TOF}
\end{center}
\end{figure}

As shown in Figure~\ref{fig:TOF}, the time of flight is more than $\sim$1.4 ns for the full length of the MS (40 cm), and more than 300 ps for one magnet sector (10 cm), therefore a resolution of 100 ps would be sufficient to determine the flight direction of the particle. Note that even for particles traversing only 1 magnet sector, 7 consecutive measurements of dE/dx will be performed, which can also be used to determine the entry direction of the particles.

The baseline design of the TOF detector is a solid plastic scintillator plate, 3 mm thick and 120$\times$120 mm$^2$ in size, readout by multiple Silicon Photomultipliers (SiPMs), with a size of 3$\times$3 mm$^2$, placed on the sides of the scintillator. A Z=1 MIP particle produces approximately 6000 scintillation photons in 3 mm of scintillator, only a fraction (a few percent) of these need to be collected for a single SiPM to allow for time and charge measurements. The SiPMs are placed equidistantly with a baseline design of 4 SiPMs per side, covering a total of 10\% of the side of the scintillator. This novel readout scheme helps to suppress the random dark noise of the SiPM by requiring coincidence of several SiPMs, and to improve the time resolution with multiple simultaneous measurements. It also allows for rough position measurements of the track using the relative counting rates of the different SiPMs. 

This flexible design provides many handles to optimize the timing and the charge measurements, since a larger SiPM surface area naturally results in a higher number of the collected photo-electrons, but decreases the intrinsic timing capabilities~\citep{SiPM}. 

\paragraph{Pixel} During the strongest solar events it is possible that pileup (more than one particle entering the same detection element) will occur with the silicon strip detector, and in particular the TOF detector, because of its relative large size. In this case the charge measurement will become challenging. The pileup effect can be eliminated by using silicon pixel detectors, which have been developed specifically for high rate environment at particle colliders up to many GHz/cm$^2$, with extremely low noise thanks to the small input capacitance. Thanks to its excellent performance, this technology has now been applied to many disciplines, including space science. As mentioned above, the Timepix hybrid pixel detector technology has been successfully applied for space radiation monitoring and space weather analyses~\citep{Granja}, as well as radiation safety in the ISS~\citep{Turecek}. 

The state-of-the-art hybrid active pixel detector is the Timepix3\footnote{\url{https://medipix.web.cern.ch/technology-chip/timepix3-chip}}
, developed within the Medipix3 collaboration. It features a cell size of 55$\mu$m$\times$55$\mu$m, which is more than sufficient to eliminate the pileup effect for the high rate events of PAN, typically not exceeding MHz/cm$^2$. A main feature of the chip is the data-driven readout scheme, keeping most of the sensor active while reading out only the pixel which has been triggered by a particle interaction. The per pixel dead time is as approximately 475 ns.

The main challenge of using large area pixel detector in space for particle detection is the power consumption. In normal operation mode Timepix3  consumes about 0.5 W/cm$^2$. To cover the surface at both ends of the spectrometer, the total power consumption will reach 80 W, much higher than the 20 W total power budget goal. Interestingly Timepix3 has a power saving mode that in principle can reduce the analogue power consumption, contributing about 50\% to the total, by a factor of 10, thus effectively reducing power consumption by half. Further power reduction on the digital side is possible by reducing the clock speed from 200 MHz to a lower value. The time stamp resolution will worsen from 1.56 ns but even a 100 ns resolution will not cause ambiguities in pixel hit assignment since the tracker can link the track to the pixel hit with high precision, and the probability that the pixel detector sees 2 particles within 100 ns is very low. A next generation chip Timepix4\footnote{\url{https://medipix.web.cern.ch/collaboration/medipix4-collaboration}}
is being developed with 65 nm CMOS technology (Timepix3 uses 130 nm CMOS). Timepix4 is 4-side buttable (Timepix3 is 3-side buttable), therefore suitable for the PAN application.

\paragraph{Thermal control} For a spectrometer it is crucial to have a stable temperature to ensure the mechanical stability of the tracking system. The thermal stability requirement is about $\pm$2$^{\circ}$C per hour. This can be achieved by relying on a passive cooling system: several heat pipes will be thermally connected to a radiator with a stable temperature ($\pm$2$^{\circ}$C/hour, at a temperature as low as possible but between $-$30$^{\circ}$C and $+$20$^{\circ}$C), installed at a suitable place on the spacecraft. A particular care will be taken to minimize the thermal resistance between the chips on PCB and the heat pipe connections. Note that thermal control systems with similar requirements have been successfully implemented on several space missions, such as PMELA, AMS-02, Fermi and DAMPE. 

A stable temperature is also important for the SiPM readout since the breakdown voltage, thus the gain of the SiPM has a temperature dependence. However within the $\pm$2$^{\circ}$C/hour limit the gain drift will not affect the detection efficiency. Longer term temperature effects can be precisely calibrated with the data itself which provides clear photo-electron peaks. Another more elaborated option would be to use temperated feedback control on the SiPM biasing voltage to ensure a temperature independent gain. 

\section{Conclusions}
The detection of energetic particles is a key component for the study of the near-Earth interplanetary space and the space-planetary environment interactions in general. From a technological point of view, the precise measurement of energetic particles in space is an extremely challenging topic. Our study shows that using the a magnetic spectrometer with permanent magnet, and employing the state-of-the-art particle detectors, such as thin and ultra precise silicon strip detectors, hybrid silicon pixel detectors and silicon photomultipliers, it is possible to develop an instrument with moderate mass and power consumption, suitable for solar and interplanetary missions, that can measure and monitor with great precision energetic particles in the 100 MeV/n to 20 GeV/n range, over a long period of time. This instrument, the Penetrating particle ANalyzer (PAN), will fill this observation gap of energetic particles in deep space, and enable ground-braking measurements that will have strong implication for cosmic ray physics, solar-terrestrial physics, circum-terrestrial and planetary space weather science and space travel.

\section*{References}

\bibliography{mybibfile_revision}

\begin{thebibliography}{36}
\expandafter\ifx\csname natexlab\endcsname\relax\def\natexlab#1{#1}\fi
\providecommand{\url}[1]{\texttt{#1}}
\providecommand{\href}[2]{#2}
\providecommand{\path}[1]{#1}
\providecommand{\DOIprefix}{doi:}
\providecommand{\ArXivprefix}{arXiv:}
\providecommand{\URLprefix}{URL: }
\providecommand{\Pubmedprefix}{pmid:}
\providecommand{\doi}[1]{\href{http://dx.doi.org/#1}{\path{#1}}}
\providecommand{\Pubmed}[1]{\href{pmid:#1}{\path{#1}}}
\providecommand{\bibinfo}[2]{#2}
\ifx\xfnm\relax \def\xfnm[#1]{\unskip,\space#1}\fi
\bibitem[{Aguilar et~al.(2018)Aguilar, Cavasonza, Alpat et~al.}]{AMS_flux}
\bibinfo{author}{Aguilar, M.}, \bibinfo{author}{Cavasonza, L.A.},
  \bibinfo{author}{Alpat, B.}, et~al., \bibinfo{year}{2018}.
\newblock \bibinfo{title}{Observation of fine time structures in the cosmic
  proton and helium fluxes with the alpha magnetic spectrometer on the
  international space station}.
\newblock \bibinfo{journal}{Phys. Rev. Lett.} \bibinfo{volume}{121},
  \bibinfo{pages}{051101}.
\bibitem[{Alexander et~al.(2016)Alexander, M.Battaglieri, B.Echenard
  et~al.}]{Alexander}
\bibinfo{author}{Alexander, J.}, \bibinfo{author}{M.Battaglieri},
  \bibinfo{author}{B.Echenard}, et~al., \bibinfo{year}{2016}.
\newblock \bibinfo{title}{Dark {S}ectors 2016 {W}orkshop: Community {R}eport.}
\newblock \bibinfo{journal}{Based on the Dark Sectors workshop at SLAC in April
  2016} \href{http://arxiv.org/abs/608.08632 (2016)}{\tt arXiv:608.08632
  (2016)}.
\bibitem[{{Aschwanden}(2002)}]{Aschwanden}
\bibinfo{author}{{Aschwanden}, M.J.}, \bibinfo{year}{2002}.
\newblock \bibinfo{title}{{Particle acceleration and kinematics in solar flares
  - A Synthesis of Recent Observations and Theoretical Concepts (Invited
  Review)}}.
\newblock \bibinfo{journal}{Space Science Reviews} \bibinfo{volume}{101},
  \bibinfo{pages}{1--227}.
\bibitem[{Ballabriga et~al.(2018)Ballabriga, Campbell and
  Llopart}]{TimepixPaper}
\bibinfo{author}{Ballabriga, R.}, \bibinfo{author}{Campbell, M.},
  \bibinfo{author}{Llopart, X.}, \bibinfo{year}{2018}.
\newblock \bibinfo{title}{Asic developments for radiation imaging applications:
  The medipix and timepix family}.
\newblock \bibinfo{journal}{Nucl. Instrum. Meth.} \bibinfo{volume}{A878},
  \bibinfo{pages}{10--23}.
\bibitem[{{Bl{\"u}mier} and Casanova(2016)}]{HAL_scheme}
\bibinfo{editor}{{Bl{\"u}mier}, P.}, \bibinfo{editor}{Casanova, F.} (Eds.),
  \bibinfo{year}{2016}.
\newblock \bibinfo{title}{Chapter 5, {M}obile {NMR} and {MRI}: {D}evelopments
  and {A}pplications ({N}ew {D}evelopments in {NMR})}.
\newblock \bibinfo{publisher}{The Royal Society of Chemistry},
  \bibinfo{address}{Cham}.
\bibitem[{Boudaud et~al.(2017)Boudaud, Lavalle and Salati}]{Boudaud}
\bibinfo{author}{Boudaud, M.}, \bibinfo{author}{Lavalle, J.},
  \bibinfo{author}{Salati, P.}, \bibinfo{year}{2017}.
\newblock \bibinfo{title}{Novel cosmic-ray electron and positron constraints on
  mev dark matter particles}.
\newblock \bibinfo{journal}{Phys. Rev. Lett.} \bibinfo{volume}{119},
  \bibinfo{pages}{021103}.
\bibitem[{{Giacalone} and {K{\'o}ta}(2006)}]{Giacalone}
\bibinfo{author}{{Giacalone}, J.}, \bibinfo{author}{{K{\'o}ta}, J.},
  \bibinfo{year}{2006}.
\newblock \bibinfo{title}{{Acceleration of Solar-Energetic Particles by
  Shocks}}.
\newblock \bibinfo{journal}{Space Science Reviews} \bibinfo{volume}{124},
  \bibinfo{pages}{277--288}.
\bibitem[{{Gluckstern}(1963)}]{Gluckstern}
\bibinfo{author}{{Gluckstern}, R.L.}, \bibinfo{year}{1963}.
\newblock \bibinfo{title}{{Uncertainties in track momentum and direction, due
  to multiple scattering and measurement errors}}.
\newblock \bibinfo{journal}{Nuclear Instruments and Methods}
  \bibinfo{volume}{24}, \bibinfo{pages}{381--389}.
\bibitem[{Gohl et~al.(2016)Gohl, Bergmann, Granja, Owens, Pichotka, Polansky
  and Pospisil}]{Gohl}
\bibinfo{author}{Gohl, S.}, \bibinfo{author}{Bergmann, B.},
  \bibinfo{author}{Granja, C.}, \bibinfo{author}{Owens, A.},
  \bibinfo{author}{Pichotka, M.}, \bibinfo{author}{Polansky, S.},
  \bibinfo{author}{Pospisil, S.}, \bibinfo{year}{2016}.
\newblock \bibinfo{title}{Measurement of particle directions in low earth orbit
  with a timepix}.
\newblock \bibinfo{journal}{Journal of Instrumentation} \bibinfo{volume}{11},
  \bibinfo{pages}{C11023}.
\bibitem[{Granja et~al.(2016)Granja, Polansky, Vykydal, Pospisil, Owens,
  Kozacek, Mellab and Simcak}]{Granja}
\bibinfo{author}{Granja, C.}, \bibinfo{author}{Polansky, S.},
  \bibinfo{author}{Vykydal, Z.}, \bibinfo{author}{Pospisil, S.},
  \bibinfo{author}{Owens, A.}, \bibinfo{author}{Kozacek, Z.},
  \bibinfo{author}{Mellab, K.}, \bibinfo{author}{Simcak, M.},
  \bibinfo{year}{2016}.
\newblock \bibinfo{title}{The {SATRAM} {T}imepix spacecraft payload in open
  space on board the {P}roba-{V} satellite for wide range radiation monitoring
  in leo orbit}.
\newblock \bibinfo{journal}{Planet. Space Sci.} \bibinfo{volume}{125},
  \bibinfo{pages}{114--129}.
\bibitem[{Halbach(1979)}]{Halbach}
\bibinfo{author}{Halbach, K.}, \bibinfo{year}{1979}.
\newblock \bibinfo{title}{Strong rare earth cobalt quadrupoles}.
\newblock \bibinfo{journal}{IEEE Transactions on Nuclear Science}
  \bibinfo{volume}{NS--26}, \bibinfo{pages}{3882--3884}.
\bibitem[{Halbach(1985)}]{Halbach1985}
\bibinfo{author}{Halbach, K.}, \bibinfo{year}{1985}.
\newblock \bibinfo{title}{Permanent magnets for production and use of high
  energy particle beams}.
\newblock \bibinfo{journal}{Lawrence Berkeley Laboratory Report, University of
  California Berkeley, California, LBL-19285 (1985)} .
\bibitem[{Laitinen(2016)}]{Laitinen}
\bibinfo{author}{Laitinen, T.}, \bibinfo{year}{2016}.
\newblock \bibinfo{title}{Experimental overview on future solar and
  heliospheric research}.
\newblock \bibinfo{journal}{XXV ECRS 2016 Proceedings - eConf C16-09-04.3}
  \href{http://arxiv.org/abs/1702.05091}{\tt arXiv:1702.05091}.
\bibitem[{Lilensten et~al.(2014)Lilensten, Coates, Dehant, de~Wit, Horne,
  Leblanc, Luhmann, Woodfield and Barthelemy}]{Lilensten}
\bibinfo{author}{Lilensten, J.}, \bibinfo{author}{Coates, A.},
  \bibinfo{author}{Dehant, V.}, \bibinfo{author}{de~Wit, T.D.},
  \bibinfo{author}{Horne, R.}, \bibinfo{author}{Leblanc, F.},
  \bibinfo{author}{Luhmann, J.}, \bibinfo{author}{Woodfield, E.},
  \bibinfo{author}{Barthelemy, M.}, \bibinfo{year}{2014}.
\newblock \bibinfo{title}{What characterizes planetary space weather?}
\newblock \bibinfo{journal}{Astron. Astrophys. Rev.} \bibinfo{volume}{22},
  \bibinfo{pages}{79}.
\bibitem[{Lubelsmeyer et~al.(2011)Lubelsmeyer, von Dratzig, Wlochala
  et~al.}]{AMS}
\bibinfo{author}{Lubelsmeyer, K.}, \bibinfo{author}{von Dratzig, A.S.},
  \bibinfo{author}{Wlochala, M.}, et~al., \bibinfo{year}{2011}.
\newblock \bibinfo{title}{Upgrade of the {A}lpha {M}agnetic {S}pectrometer
  ({AMS}-02) for long term operation on the {I}nternational {S}pace {S}tation
  ({ISS})}.
\newblock \bibinfo{journal}{Nucl. Instrum. Meth.} \bibinfo{volume}{A654},
  \bibinfo{pages}{639--648}.
\bibitem[{Malandraki and Crosby(2018)}]{Malandraki}
\bibinfo{editor}{Malandraki, O.E.}, \bibinfo{editor}{Crosby, N.B.} (Eds.),
  \bibinfo{year}{2018}.
\newblock \bibinfo{title}{Solar Particle Radiation Storms Forecasting and
  Analysis - The HESPERIA HORIZON 2020 Project and Beyond}.
\newblock \bibinfo{publisher}{Springer}, \bibinfo{address}{Cham}.
\bibitem[{Martucci et~al.(2018)Martucci, Munini, Boezio et~al.}]{PAM_flux}
\bibinfo{author}{Martucci, M.}, \bibinfo{author}{Munini, R.},
  \bibinfo{author}{Boezio, M.}, et~al., \bibinfo{year}{2018}.
\newblock \bibinfo{title}{Proton fluxes measured by the pamela experiment from
  the minimum to the maximum solar activity for solar cycle 24}.
\newblock \bibinfo{journal}{The Astrophysical Journal Letters}
  \bibinfo{volume}{854}, \bibinfo{pages}{L2}.
\bibitem[{{M{\"u}ller-Mellin} et~al.(1995){M{\"u}ller-Mellin}, Kunow,
  Flei{\ss}ner et~al.}]{Mellin}
\bibinfo{author}{{M{\"u}ller-Mellin}, R.}, \bibinfo{author}{Kunow, H.},
  \bibinfo{author}{Flei{\ss}ner, V.}, et~al., \bibinfo{year}{1995}.
\newblock \bibinfo{title}{{COSTEP - Comprehensive Suprathermal and Energetic
  Particle Analyser}}.
\newblock \bibinfo{journal}{Solar Physics} \bibinfo{volume}{162},
  \bibinfo{pages}{483--504}.
\bibitem[{Patrignani and {Particle Data Group}(2016)}]{PDG}
\bibinfo{author}{Patrignani, C.}, \bibinfo{author}{{Particle Data Group}},
  \bibinfo{year}{2016}.
\newblock \bibinfo{title}{Review of particle physics}.
\newblock \bibinfo{journal}{Chinese Physics C} \bibinfo{volume}{40},
  \bibinfo{pages}{100001}.
\bibitem[{Picozza et~al.(2007)Picozza, Galper, Castellini et~al.}]{PAM}
\bibinfo{author}{Picozza, P.}, \bibinfo{author}{Galper, A.M.},
  \bibinfo{author}{Castellini, G.}, et~al., \bibinfo{year}{2007}.
\newblock \bibinfo{title}{{PAMELA} - {A} {P}ayload for {A}ntimatter {M}atter
  {E}xploration and {L}ight-nuclei {A}strophysics}.
\newblock \bibinfo{journal}{Astroparticle Physics} \bibinfo{volume}{27},
  \bibinfo{pages}{296--315}.
\bibitem[{Plainaki et~al.(2007)Plainaki, Belov, Eroshenko, Mavromichalaki and
  Yanke}]{Plainaki2007}
\bibinfo{author}{Plainaki, C.}, \bibinfo{author}{Belov, A.},
  \bibinfo{author}{Eroshenko, E.}, \bibinfo{author}{Mavromichalaki, H.},
  \bibinfo{author}{Yanke, V.}, \bibinfo{year}{2007}.
\newblock \bibinfo{title}{Modeling ground level enhancements: Event of 20
  {J}anuary 2005}.
\newblock \bibinfo{journal}{Journal of Geophysical Research (Space Physics)}
  \bibinfo{volume}{112}, \bibinfo{pages}{A04102}.
\bibitem[{Plainaki et~al.(2016)Plainaki, J.Lilensten, Radioti
  et~al.}]{Plainaki2016}
\bibinfo{author}{Plainaki, C.}, \bibinfo{author}{J.Lilensten},
  \bibinfo{author}{Radioti, A.}, et~al., \bibinfo{year}{2016}.
\newblock \bibinfo{title}{Planetary space weather: Scientific aspects and
  future perspectives}.
\newblock \bibinfo{journal}{Journal of Space Weather and Space Climate}
  \bibinfo{volume}{6}, \bibinfo{pages}{A31}.
\bibitem[{Plainaki et~al.(2014)Plainaki, Mavromichalaki, Laurenza
  et~al.}]{Plainaki}
\bibinfo{author}{Plainaki, C.}, \bibinfo{author}{Mavromichalaki, H.},
  \bibinfo{author}{Laurenza, M.}, et~al., \bibinfo{year}{2014}.
\newblock \bibinfo{title}{The ground-level enhancement of 2012 may 17:
  Derivation of solar proton event properties through the application of the
  nmbangle ppola model}.
\newblock \bibinfo{journal}{The Astrophysical Journal} \bibinfo{volume}{785},
  \bibinfo{pages}{160}.
\bibitem[{Potgieter(2013)}]{Potgieter2}
\bibinfo{author}{Potgieter, M.S.}, \bibinfo{year}{2013}.
\newblock \bibinfo{title}{Solar modulation of cosmic rays}.
\newblock \bibinfo{journal}{Living Reviews in Solar Physics}
  \bibinfo{volume}{10}, \bibinfo{pages}{3}.
\bibitem[{{Potgieter} and {Vos}(2017)}]{Potgieter1}
\bibinfo{author}{{Potgieter}, M.S.}, \bibinfo{author}{{Vos}, E.E.},
  \bibinfo{year}{2017}.
\newblock \bibinfo{title}{{Difference in the heliospheric modulation of
  cosmic-ray protons and electrons during the solar minimum period of 2006 to
  2009}}.
\newblock \bibinfo{journal}{Astronomy \& Astrophysics} \bibinfo{volume}{601},
  \bibinfo{pages}{A23}.
\bibitem[{Puill et~al.(2012)Puill, C.Bazin, D.Breton et~al.}]{SiPM}
\bibinfo{author}{Puill, V.}, \bibinfo{author}{C.Bazin},
  \bibinfo{author}{D.Breton}, et~al., \bibinfo{year}{2012}.
\newblock \bibinfo{title}{Single photoelectron timing resolution of {SiPM} as a
  function of the bias voltage, the wavelength and the temperature}.
\newblock \bibinfo{journal}{Nucl. Instrum. Meth.} \bibinfo{volume}{A695},
  \bibinfo{pages}{354--358}.
\bibitem[{{Reames}(1999)}]{Reames}
\bibinfo{author}{{Reames}, D.V.}, \bibinfo{year}{1999}.
\newblock \bibinfo{title}{{Acceleration of Solar-Energetic Particles by
  Shocks}}.
\newblock \bibinfo{journal}{Space Science Reviews} \bibinfo{volume}{90},
  \bibinfo{pages}{413--491}.
\bibitem[{{Reames}(2013)}]{Reames2013}
\bibinfo{author}{{Reames}, D.V.}, \bibinfo{year}{2013}.
\newblock \bibinfo{title}{{The Two Sources of Solar Energetic Particles}}.
\newblock \bibinfo{journal}{Space Science Reviews} \bibinfo{volume}{175},
  \bibinfo{pages}{53--92}.
\bibitem[{Soltner and {Bl{\"u}mier}(2010)}]{Soltner}
\bibinfo{author}{Soltner, H.}, \bibinfo{author}{{Bl{\"u}mier}, P.},
  \bibinfo{year}{2010}.
\newblock \bibinfo{title}{Dipolar {H}albach magnet stacks made from identically
  shaped permanent magnets for magnetic resonance}.
\newblock \bibinfo{journal}{Concepts Magn. Reson. Part A}
  \bibinfo{volume}{36A}, \bibinfo{pages}{211--222}.
\bibitem[{Stone et~al.(1998)Stone, Cohen, Cook et~al.}]{ACE}
\bibinfo{author}{Stone, E.C.}, \bibinfo{author}{Cohen, C.},
  \bibinfo{author}{Cook, W.}, et~al., \bibinfo{year}{1998}.
\newblock \bibinfo{title}{The cosmic-ray isotope spectrometer for the
  {A}dvanced {C}omposition {E}xplorer}.
\newblock \bibinfo{journal}{Space Science Reviews} \bibinfo{volume}{96},
  \bibinfo{pages}{285--356}.
\bibitem[{Stone et~al.(2013)Stone, Cummings, McDonald et~al.}]{Voyager}
\bibinfo{author}{Stone, E.C.}, \bibinfo{author}{Cummings, A.C.},
  \bibinfo{author}{McDonald, F.B.}, et~al., \bibinfo{year}{2013}.
\newblock \bibinfo{title}{Voyager 1 observes low-energy galactic cosmic rays in
  a region depleted of heliospheric ions}.
\newblock \bibinfo{journal}{Science} \bibinfo{volume}{341},
  \bibinfo{pages}{150--153}.
\bibitem[{Straulino(2006)}]{PAMELA_tracker}
\bibinfo{author}{Straulino, S.}, \bibinfo{year}{2006}.
\newblock \bibinfo{title}{Spatial resolution of double-sided silicon microstrip
  detectors for the {PAMELA} apparatus}.
\newblock \bibinfo{journal}{Nucl. Instrum. Meth.} \bibinfo{volume}{A556},
  \bibinfo{pages}{100--114}.
\bibitem[{Thonet(2016)}]{Thonet}
\bibinfo{author}{Thonet, P.A.}, \bibinfo{year}{2016}.
\newblock \bibinfo{title}{Use of permanent magnets in multiple projects at
  cern}.
\newblock \bibinfo{journal}{IEEE Trans on Appl. Superconductivity}
  \bibinfo{volume}{26}, \bibinfo{pages}{1--4}.
\bibitem[{Turchetta(1993)}]{charge}
\bibinfo{author}{Turchetta, R.}, \bibinfo{year}{1993}.
\newblock \bibinfo{title}{Spatial resolution of silicon microstrip detectors}.
\newblock \bibinfo{journal}{Nucl. Instrum. Meth.} \bibinfo{volume}{A335},
  \bibinfo{pages}{44--58}.
\bibitem[{Turecek et~al.(2011)Turecek, Pinsky, Jakubek, Vykydal, Stoffle and
  Pospisil}]{Turecek}
\bibinfo{author}{Turecek, D.}, \bibinfo{author}{Pinsky, L.},
  \bibinfo{author}{Jakubek, J.}, \bibinfo{author}{Vykydal, Z.},
  \bibinfo{author}{Stoffle, N.}, \bibinfo{author}{Pospisil, S.},
  \bibinfo{year}{2011}.
\newblock \bibinfo{title}{Small dosimeter based on timepix device for
  international space station}.
\newblock \bibinfo{journal}{Journal of Instrumentation} \bibinfo{volume}{6},
  \bibinfo{pages}{C12037}.
\bibitem[{Xapsos et~al.(2012)Xapsos, Stauffer, Jordan et~al.}]{Xapsos}
\bibinfo{author}{Xapsos, M.A.}, \bibinfo{author}{Stauffer, C.A.},
  \bibinfo{author}{Jordan, T.M.}, et~al., \bibinfo{year}{2012}.
\newblock \bibinfo{title}{Periods of high intensity solar proton flux}.
\newblock \bibinfo{journal}{IEEE Transactions on Nuclear Science}
  \bibinfo{volume}{59}, \bibinfo{pages}{1054--1059}.

\end{thebibliography}

\end{document}